\documentclass[aps,showpacs,tightenlines,twocolumn,nofootinbib,nobibnotes,superscriptaddress]{revtex4-1}
\usepackage{amsmath,amssymb,amsfonts,bm}
\usepackage{graphicx}
\usepackage{epstopdf}
\usepackage{dcolumn}
\usepackage{mathrsfs}
\usepackage{tgtermes}
\usepackage[colorlinks=true,linkcolor=red,citecolor=blue, urlcolor=blue,bookmarks=false]{hyperref}
\bibpunct{[}{]}{,}{n}{}{}
\usepackage{verbatim}
\usepackage{amsthm,amsmath,amssymb}
\usepackage{mathrsfs}
\usepackage{booktabs}
\usepackage{multirow}

\begin{document}
\title{Tunable cornerlike states in topological type-II hyperbolic lattices}
\date{\today}
\author{Zheng-Rong Liu}
\affiliation{Department of Physics, Hubei University, Wuhan 430062, China}
\author{Tan Peng}
\affiliation{Collaborative Innovation Center for Optoelectronic Technology of Ministry of Education and Hubei Province (Hubei University of Automotive Technology), and Shiyan Key Laboratory of Quantum Information and Precision Optics, Shiyan 442002, China}
\author{Xiang Liu}
\affiliation{Department of Physics, Hubei University, Wuhan 430062, China}
\author{Xiao-Xia Yi}
\affiliation{Department of Fundamental Subjects, Wuchang Shouyi University, Wuhan 430064, China}
\author{Chun-Bo Hua}
\affiliation{School of Electronic and Information Engineering, Hubei University of Science and Technology, Xianning 437100, China}
\affiliation{Key Laboratory of Artificial Micro- and Nanostructures of Ministry of Education and School of Physics and Technology, Wuhan University, Wuhan 430072, China}
\author{Rui Chen}\email{chenr@hubu.edu.cn}
\affiliation{Department of Physics, Hubei University, Wuhan 430062, China}
\author{Bin Zhou}\email{binzhou@hubu.edu.cn}
\affiliation{Department of Physics, Hubei University, Wuhan 430062, China}
\affiliation{Key Laboratory of Intelligent Sensing System and Security of Ministry of Education, Hubei University, Wuhan 430062, China}
\affiliation{Wuhan Institute of Quantum Technology, Wuhan 430206, China}

\begin{abstract}
Type-II hyperbolic lattices constitute a new class of hyperbolic structures that are projected onto the Poincar\'{e} ring and possess both an inner and an outer boundary. In this work, we reveal the higher-order topological phases in type-II hyperbolic lattices, characterized by the generalized quadrupole moment. Unlike the type-I hyperbolic lattices where zero-energy cornerlike states exist on a single boundary, the higher-order topological phases in type-II hyperbolic lattices possess zero-energy cornerlike states localized on both the inner and outer boundaries. These findings are verified within both the modified Bernevig-Hughes-Zhang model and the Benalcazar-Bernevig-Hughes model. Furthermore, we demonstrate that the higher-order topological phase remains robust against weak disorder in type-II hyperbolic lattices. Our work provides a route for realizing and controlling higher-order topological states in type-II hyperbolic lattices.
\end{abstract}

\maketitle

\section{Introduction}
The extension of topological phases of matter from flat Euclidean geometries to curved non-Euclidean spaces has recently emerged as a rapidly growing frontier~\cite{PhysRevLett.125.053901, Kollar2020CMP, PhysRevA.102.032208, doi:10.1126/sciadv.abe9170, Ikeda_2021, RUZZENE2021101491, Zhu_2021, doi:10.1073/pnas.2116869119, PhysRevLett.129.088002, PhysRevB.105.125118, PhysRevE.106.034114, Zhang2022, PhysRevB.105.245301, PhysRevLett.129.246402, PhysRevB.106.155146, PhysRevLett.128.166402, PhysRevB.106.155120, Lenggenhager2022, PhysRevLett.128.013601, 10.1038/s41467-023-36359-6, 10.1038/s41467-023-36767-8, PhysRevB.107.125302, PhysRevB.107.184201, 10.1038/s41467-024-46035-y, 10.1038/s41467-024-46551-x, PhysRevLett.132.206601}. Hyperbolic lattices, which realize regular tessellations in spaces of constant negative curvature, have proven to be a particularly fertile platform, enabling theoretical and experimental advances ranging from hyperbolic quantum spin Hall states~\cite{PhysRevLett.125.053901, PhysRevLett.129.246402} and Chern insulators~\cite{Zhang2022, PhysRevB.105.245301, PhysRevLett.129.246402} to higher-order hyperbolic topological phases~\cite{PhysRevB.107.125302, PhysRevB.107.184201}. Conventional type-I hyperbolic lattices, mapped from two-sheet hyperboloids~\cite{Zhang2022, PhysRevB.105.125118, 10.1038/s42005-025-01990-w}, host only a single boundary, whereas the recently introduced type-II hyperbolic lattices, derived from one-sheet hyperboloids, possess both inner and outer boundaries~\cite{arXiv2305.04862, PhysRevLett.133.061603, 10.1038/s42005-025-01990-w}. This fundamental geometric distinction not only enriches the boundary structures available in non-Euclidean systems but also endows type-II hyperbolic lattices with the capacity to support boundary phenomena absent in type-I counterparts. Recent studies have revealed that type-II hyperbolic lattices sustain unconventional Chern insulating phases with counterpropagating edge channels and enable dynamical processes that transfer topological states across distinct edges~\cite{10.1038/s42005-025-01990-w}. These developments set the stage for investigating higher-order topological insulators (HOTIs) in type-II hyperbolic geometries.

HOTIs have emerged as a natural generalization of conventional topological phases, where protected boundary states appear in dimensions lower than expected from the usual bulk--boundary correspondence, such as zero-dimensional corner modes in two-dimensional systems or one-dimensional hinge states in three dimensional systems~\cite{doi:10.1126/science.aah6442, PhysRevLett.119.246401, PhysRevLett.119.246402, PhysRevB.96.245115, PhysRevLett.120.026801, Serra_Garcia_2018, doi:10.1126/sciadv.aat0346, Peterson2018, PhysRevB.97.155305, PhysRevB.97.205135, PhysRevB.97.205136, Noh2018, Schindler2018, PhysRevX.8.031070, PhysRevB.98.201114, 10.1038/s41567-018-0246-1, PhysRevX.9.011012, PhysRevLett.122.076801, 10.1038/s41563-018-0251-x, 10.1038/s41563-018-0252-9, PhysRevLett.122.256402, 10.1038/s41567-019-0457-0, PhysRevB.99.245151, PhysRevLett.123.256402, PhysRevB.100.235302, PhysRevLett.124.036803, 10.1038/s41535-019-0206-8, PhysRevLett.124.136407, PhysRevLett.124.166804, PhysRevB.102.241102, PhysRevLett.127.066801, PhysRevLett.127.196801, PhysRevB.104.245302, PhysRevB.92.085126, https://doi.org/10.1002/lpor.202200499, 10.1088/1674-1056/ada885}. These exotic boundary excitations have been experimentally observed across diverse platforms, including photonic crystals~\cite{Noh2018, 10.1038/s41377-020-0331-y, https://doi.org/10.1002/lpor.202200499, 10.1038/s41467-022-33894-6, 10.1038/s41566-019-0452-0}, acoustic metamaterials~\cite{Serra_Garcia_2018, PhysRevLett.122.244301, 10.1038/s41563-018-0251-x, 10.1038/s41563-018-0252-9, PhysRevLett.124.206601, 10.3389/fphy.2021.770589, PhysRevLett.132.066601}, and electric circuits~\cite{10.1038/s41567-018-0246-1, 10.1038/s42005-021-00610-7}. Beyond Euclidean systems, recent advances have revealed that higher-order topological states can also exist in hyperbolic geometries~\cite{PhysRevB.107.125302, PhysRevB.107.184201}. To date, higher-order topological phases have been extensively studied in crystalline systems~\cite{doi:10.1126/science.aah6442, PhysRevLett.119.246401, PhysRevLett.119.246402, PhysRevB.96.245115, PhysRevLett.120.026801, Serra_Garcia_2018, doi:10.1126/sciadv.aat0346, Peterson2018, PhysRevB.97.155305, PhysRevB.97.205135, PhysRevB.97.205136, Noh2018, Schindler2018, PhysRevX.8.031070, PhysRevB.98.201114, 10.1038/s41567-018-0246-1, PhysRevX.9.011012, PhysRevLett.122.076801, 10.1038/s41563-018-0251-x, 10.1038/s41563-018-0252-9, PhysRevLett.122.256402, 10.1038/s41567-019-0457-0, PhysRevB.99.245151, PhysRevLett.123.256402, PhysRevB.100.235302, 10.1038/s41535-019-0206-8, PhysRevLett.124.136407, PhysRevLett.124.166804, PhysRevLett.127.066801, PhysRevLett.127.196801, PhysRevB.92.085126, https://doi.org/10.1002/lpor.202200499}, quasicrystals~\cite{PhysRevLett.124.036803, PhysRevB.102.241102, PhysRevB.104.245302, PhysRevB.108.195306, PhysRevResearch.2.033071, PhysRevLett.123.196401}, fractal lattices~\cite{PhysRevB.105.L201301, 10.1016/j.scib.2022.09.020}, amorphous lattices~\cite{PhysRevResearch.2.012067, PhysRevLett.126.206404, PhysRevB.106.125310, PhysRevB.109.195301, PhysRevB.111.075306}, and type-I hyperbolic lattices~\cite{PhysRevB.107.125302, PhysRevB.107.184201}. However, higher-order topological phases in type-II hyperbolic lattices have not yet been reported.

\begin{figure}[t]
	\includegraphics[width=0.3\textwidth]{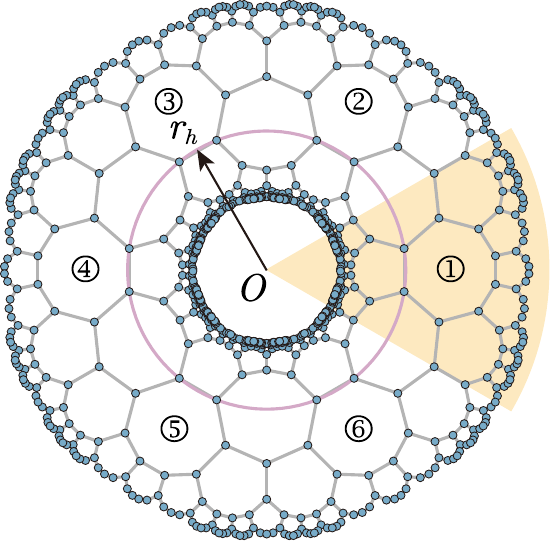} \caption{In the Poincar\'{e} ring, the vertices of the polygons correspond to the sites of a type-II hyperbolic lattice. The symbol $\{p=8, q=3, k=6\}$ denotes a tiling by regular $p$-sided polygons on the Poincar\'{e} ring, where $q$ polygons meet at each vertex in the bulk. Here, $r_{h}=e^{-2\pi/kP}$ represents the characteristic radius of the type-II hyperbolic lattice, where the structural parameter $k=6$ is an integer and $P=1.559$ is a geometry constant~\cite{arXiv2305.04862, PhysRevLett.133.061603, 10.1038/s42005-025-01990-w}. The parameter $k$ denotes the number of repeating units (yellow sector) in the lattice. A rotation by $2\pi/k$ of the sites within one unit brings them into coincidence with the sites of an adjacent repeating unit.}%
	\label{fig1}
\end{figure}

In this work, we reveal the HOTI phases in type-II hyperbolic lattices based on the modified Bernevig-Hughes-Zhang (BHZ) model~\cite{PhysRevLett.124.036803, PhysRevResearch.2.012067} and the Benalcazar-Bernevig-Hughes (BBH) model~\cite{doi:10.1126/science.aah6442}. In contrast to type-I hyperbolic lattices, the HOTI phases in type-II hyperbolic lattices feature zero-energy cornerlike states on both the inner and outer boundaries. We demonstrate that the HOTI phases with zero-energy cornerlike states are characterized by a nonzero (generalized) quadrupole moment. Meanwhile, these HOTI phases are also shown to be robust against weak disorder. In the modified BHZ model, the number of zero-energy cornerlike states is governed by the variation period of the Wilson mass term, while their spatial locations are controlled by the polarization angle of the Hamiltonian. We further reveal that the finite-size effects can be suppressed by increasing the structural parameter $k$, which is a key structural feature of type-II hyperbolic lattices. In the BBH model, the type-II hyperbolic lattices can transition from a trivial insulator to the HOTI with a nonzero quadrupole moment by tuning the system parameter.

The rest of the paper is organized as follows. In Sec.~\ref{SecIIA}, we introduce the modified BHZ model in the type-II hyperbolic lattices, and demonstrate the method for calculating the quadrupole moment in Sec.~\ref{SecIIB}. In Sec.~\ref{SecIIC}, we verify the tunability of the number of zero-energy cornerlike states in type-II hyperbolic lattices. In Sec.~\ref{SecIID}, we demonstrate that the spatial locations of the zero-energy cornerlike states can be controllably tuned. In Sec.~\ref{SecIII}, we present the phase transitions of the BBH model on type-II hyperbolic lattices. Finally, we summarize our conclusions in Sec.~\ref{Conclusion}.

\vspace{-0.1cm}
\section{Modified Bernevig-Hughes-Zhang Model in type-II hyperbolic lattices}
\label{SecII}
In this section, we investigate the phase transition of HOTIs in type-II hyperbolic lattices. Unlike conventional hyperbolic lattices, the lattice sites of a type-II hyperbolic lattice are distributed over the Poincar\'{e} ring. In Fig.~\ref{fig1}, the vertices of the polygons in the Poincar\'{e} ring, where the Poincar\'{e} ring is obtained from a one-sheeted hyperboloid by stereographic projection~\cite{10.1038/s42005-025-01990-w}, denote the sites of a type-II hyperbolic lattice, and the gray solid lines indicate the geodesics corresponding to the shortest hyperbolic distances between pairs of sites. The structure of the type-II hyperbolic lattice is characterized by the symbol $\{ p, q, k\}$, where $p=8$ denotes the number of vertices of the regular polygons tessellating the Poincar\'{e} ring, $q=3$ specifies the number of polygons meeting at each vertex in the bulk, and the structural parameter $k=6$ is an integer representing the number of repeated cells. $r_{h}=e^{-2\pi/kP}$ defines the characteristic radius of the type-II hyperbolic lattice~\cite{10.1038/s42005-025-01990-w}. The circle of radius $r_{h}$ (marked by purple circle) divides the type-II hyperbolic lattice into the inner-ring and the outer-ring. Taking this circle as the inversion circle with center $O$, sites in the inner region are mapped to those in the outer region via circle inversion, and vice versa.

\subsection{Model}
\label{SecIIA}
Here, we apply the modified BHZ model to the type-II hyperbolic lattices. The modified BHZ model can be described by~\cite{PhysRevLett.124.036803, PhysRevResearch.2.012067}
\begin{align}
H=&-\frac{1}{2}\sum_{\left <m,n \right >}c_{m}^{\dagger}it_{1}\left [s_{z}\tau_{x}\cos(\theta_{mn})+s_{0}\tau_{y}\sin(\theta_{mn})\right ]c_{n}\nonumber \\
&-\frac{1}{2}\sum_{\left <m,n \right >}c_{m}^{\dagger}t_{2}s_{0}\tau_{z}c_{n}+\sum_{m}(M+2t_{2})c_{m}^{\dagger}s_{0}\tau_{z}c_{m}\nonumber \\
&+\frac{g}{2}\sum_{\left <m,n \right >}c_{m}^{\dagger}\cos(\eta\theta_{mn})s_{x}\tau_{x}c_{n},
\label{eq1}
\end{align}
where $c_{m}^{\dagger}$ and $c_{m}$ are the creation and annihilation operators of electrons on site $m$. $\theta_{mn}$ represents the polar angle of the vector from site $n$ to site $m$ in the Poincar\'{e} ring. $s_{0}$ is the identity matrix, $s_{x,y,z}$ and $\tau_{x,y,z}$ are the Pauli matrices representing spin and orbital, respectively. $M$ denotes the Dirac mass, $t_{1}$ is the spin-orbit coupling strength, and $t_{2}$ is the hopping amplitude. In Eq.~(\ref{eq1}), the modified BHZ model consists of two components. The first three terms constitute the conventional BHZ model expressed in polar coordinates. The last term in Eq.~(\ref{eq1}) represents the Wilson mass term, and $g$ is the magnitude of the Wilson mass. $\eta$ can only take even numbers, which is used to adjust the variation period of the Wilson mass. It is worth noting that Hamiltonian $H$ does not possess translation symmetry in type-II hyperbolic lattices.

When the modified BHZ model with $\eta=2$ is applied to a square lattice, the Hamiltonian $H$ can be transformed via Fourier transformation into the following form:
\begin{align}
H_{\rm{mBHZ}}=&t_{1}\sin(k_{x})s_{z}\tau_{x}+t_{1}\sin(k_{y})s_{0}\tau_{y}\nonumber\\
&+[M+2t_{2}-t_{2}\cos(k_{x})-t_{2}\cos(k_{y})]s_{0}\tau_{z}\nonumber\\
&+g[\cos(k_{x})-\cos(k_{y})]s_{x}\tau_{x},
\label{mBHZk}
\end{align}
where the first three terms corresponds to the conventional BHZ model, and the last term represents the Wilson mass term in the square lattice for $\eta = 2$.

\begin{figure}[t]
	\includegraphics[width=0.48\textwidth]{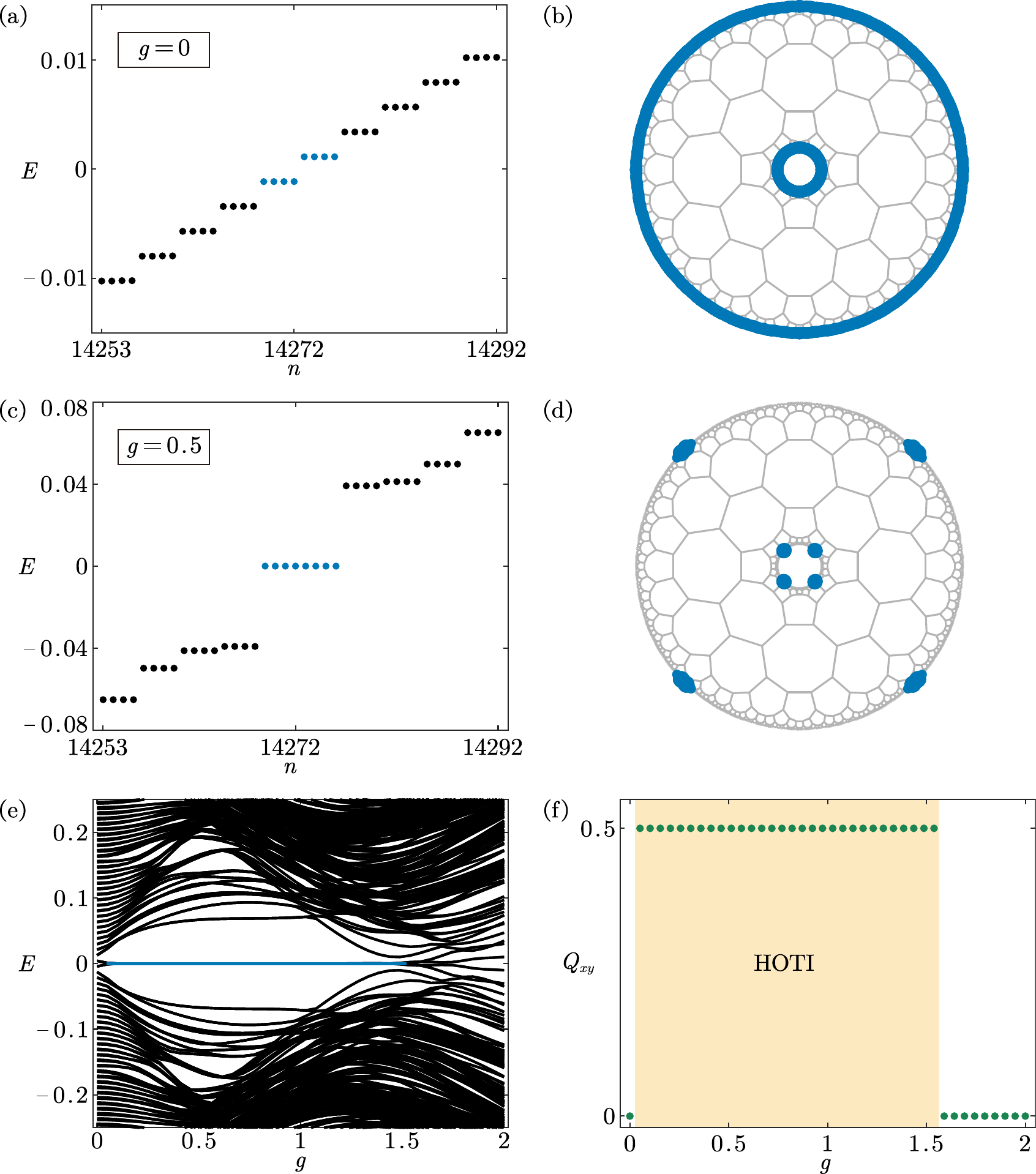} \caption{(a) Energy spectrum of the Hamiltonian $H$ in the $\{8, 3, 4\}$ lattice when $g=0$. (b) The probability distribution of the boundary states marked with blue dots in (a). (c) Energy spectrum of the Hamiltonian $H$ in the $\{8, 3, 4\}$ lattice when $g=0.5$. (d) The probability distribution of the eight zero-energy eigenstates marked with blue dots in (c). (e) Energy of the Hamiltonian $H$ as a function of the Wilson mass $g$. (f) The quadrupole moment $Q_{xy}$ as a function of the Wilson mass $g$. Here, we take the parameters $M=-1$, $t_{1}=t_{2}=1$, and $\eta=2$.}%
	\label{fig2}
\end{figure}

\begin{figure}[t]
	\includegraphics[width=0.48\textwidth]{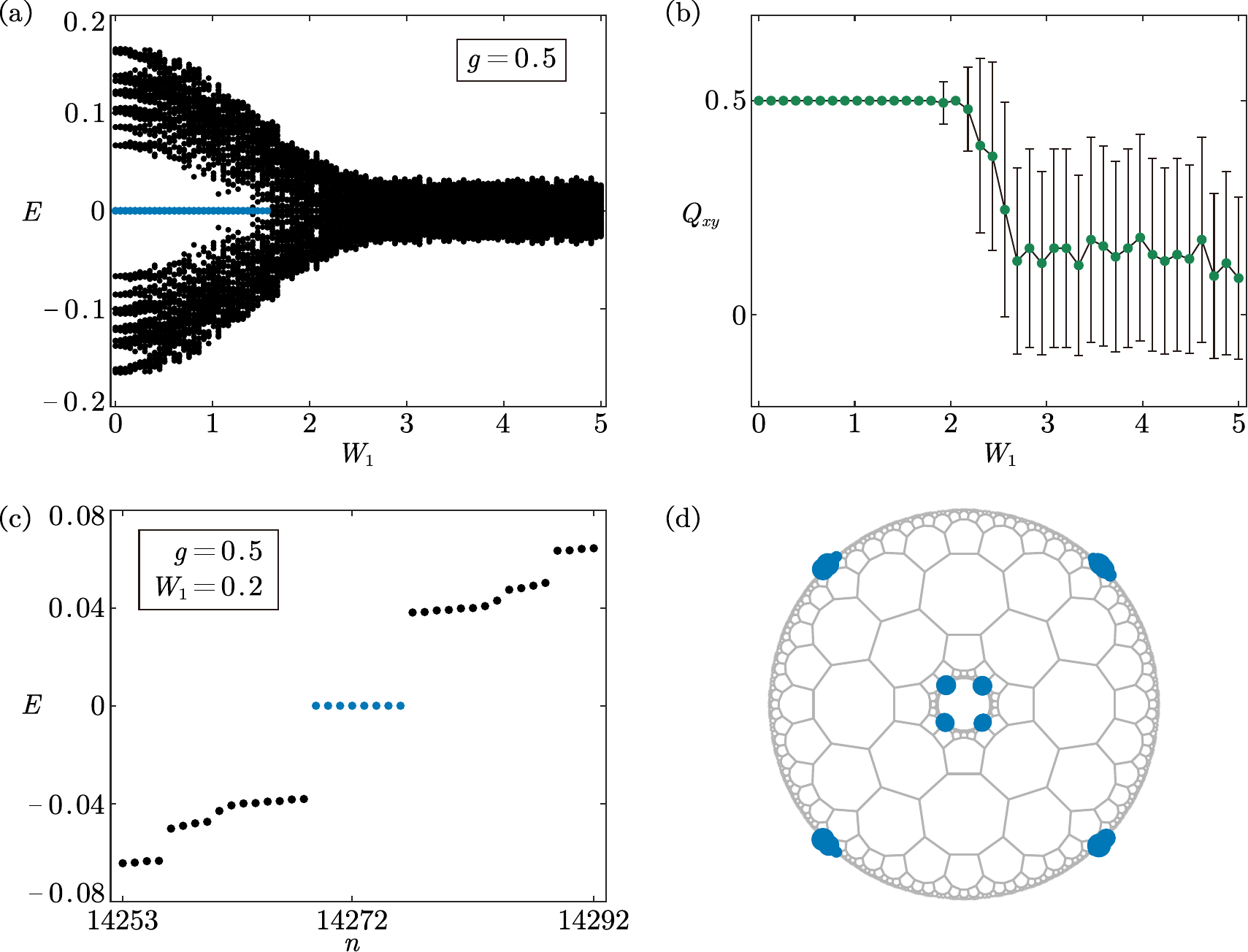} \caption{(a) Energy of the Hamiltonian $H+H_{W1}$ as a function of the disorder strength $W_{1}$ when $g=0.5$. (b) The quadrupole moment $Q_{xy}$ as a function of the disorder strength $W_{1}$ when $g=0.5$. The error bar represents the standard deviation of 100 samples. (c) Energy spectrum of the Hamiltonian $H+H_{W1}$ when $g=0.5$ and $W_{1}=0.2$. (d) The probability distribution of the eight zero-energy eigenstates marked with blue dots in (c). Here, we take the parameters $M=-1$, $t_{1}=t_{2}=1$, $\eta=2$, and $k=4$.}%
	\label{fig3}
\end{figure}

First, we investigate the phase transition of the system at $\eta=2$. Figure~\ref{fig2}(a) shows the energy spectrum of the Hamiltonian $H$ when $g=0$, where a gapless spectrum can be observed. The Poincar\'{e} ring has both inner and outer boundaries, and the Hamiltonian $H$ possesses the time-reversal symmetry $T=is_{y}\tau_{0}\mathcal{K}$ (where $\mathcal{K}$ is the complex conjugation) when $g=0$. As a result, the spectrum exhibits fourfold degeneracy. Figure~\ref{fig2}(b) shows the probability distributions of the eigenstates marked by blue dots in Fig.~\ref{fig2}(a). It is clear that these states are uniformly distributed along the inner and outer boundaries of the Poincar\'{e} ring. When the Wilson mass term is turned on, the time-reversal symmetry $T$ is broken and the boundary states become gapped. The factor $\cos(\eta\theta_{mn})$ divides each boundary into $2\eta$ segments where the Wilson mass alternates between positive and negative values. At the interfaces between these segments, zero-energy cornerlike states with vanishing Wilson mass emerge. Since both the inner and outer boundaries of the Poincar\'{e} ring host $2\eta$ such zero-energy states, a total of $4\eta$ zero-energy modes appear within the gap of the energy spectrum, as shown in Fig.~\ref{fig2}(c). Figure~\ref{fig2}(d) displays the probability distributions of these eight zero-energy states. These zero-energy cornerlike states are protected by particle-hole symmetry $P=s_{z}\tau_{x}\mathcal{K}$ and the composite symmetry $Sm_{z}$, where $S=PT$ is the chiral symmetry operator and $m_{z}=s_{z}\tau_{0}$ represents the mirror symmetry operator. In addition, we investigate the effect of the Wilson mass strength on the phase transitions in type-II hyperbolic lattices. Figure~\ref{fig2}(e) displays the evolution of the energy spectrum with respect to the Wilson mass $g$. A suitably chosen Wilson-mass term opens the boundary-state gap and gives rise to zero-energy cornerlike states. These states are eliminated when the Wilson-mass term becomes excessively strong.

\subsection{The quadrupole moment}
\label{SecIIB}
In Euclidean systems, the quadrupole moment is widely employed to characterize the topological properties of HOTIs~\cite{PhysRevB.100.245134, PhysRevB.100.245135, doi:10.1126/science.aah6442, PhysRevB.96.245115, PhysRevLett.125.166801, PhysRevB.101.195309, PhysRevResearch.2.012067, PhysRevB.104.245302, PhysRevB.103.085408}. Moreover, in previous work, we have shown that the real-space quadrupole moment can also capture higher-order topology in type-I hyperbolic lattices~\cite{PhysRevB.107.125302}. Motivated by these results, we introduce the quadrupole moment to identify higher-order topological phases in type-II hyperbolic lattices. The real-space quadrupole moment $Q_{xy}$ is given by~\cite{PhysRevLett.125.166801, PhysRevB.100.245134, PhysRevB.100.245135, PhysRevB.101.195309, PhysRevResearch.2.012067, PhysRevB.104.245302, PhysRevB.103.085408}
\begin{align}
Q_{xy}&=\left [ \frac{1}{2\pi}{\rm{Im}}~{\rm{log}}~{\rm{det}}(\Psi_{\rm occ}^{\dagger}\hat{U}\Psi_{\rm occ})-Q_{0} \right ]~{\rm{mod}}~1,
\label{eqQxy1}
\end{align}
with
\begin{align}
Q_{0}&=\frac{1}{2}\sum_{j}x_{m}y_{m}/A_{\rm{HL}},
\label{eqQxy2}
\end{align}
where $\Psi_{\rm occ}$ are the occupied eigenstates of $H$, $\hat{U}$ is a diagonal matrix whose diagonal elements are $e^{2\pi i x_{m}y_{m}/A_{\rm{HL}}}$, and $(x_{m},y_{m})$ denotes the rescaled coordinate of the $m$th site in the Poincar\'{e} ring. $A_{\rm{HL}}=\pi r_{\rm{out}}^{2}-\pi r_{\rm{in}}^{2}$ is the area of the Poincar\'{e} ring with the outer radius $r_{\rm{out}}=1$ and the inner radius $r_{\rm{in}}=r_{h}^{2}=e^{-4\pi/kP}$. In calculations, we need to translate the coordinates in interval $x_{m},y_{m}\in(-1,1)$ to interval $x_{m},y_{m}\in(0,2)$. The HOTI phase is characterized by the quadrupole moment $Q_{xy}=0.5$, while the quadrupole moment of a trivial system is equal to $0$. In Fig.~\ref{fig2}(f), we present the evolution of the quadrupole moment $Q_{xy}$ as a function of the Wilson mass $g$. At $g=0$, the system resides in a quantum spin Hall insulator phase. When a suitably Wilson mass term is introduced, the boundary-state gap is opened and the system transitions into a HOTI phase characterized by a nonzero quadrupole moment of $Q_{xy}=0.5$. When the Wilson mass exceeds a critical value $g\approx1.55$, the zero-energy states are eliminated, and the quadrupole moment $Q_{xy}$ drops to zero. It is worth noting that both the radial expansion and the coordination number $q$ in type-II hyperbolic lattices influence the topological phase transition (see Appendices~\ref{AppendixA} and \ref{AppendixB} for more details).

Furthermore, to verify the robustness of the zero-energy localized states against disorder, we introduce the on-site disorder $H_{W1}=W_{1}\sum_{m}c^{\dagger}_{m}\omega_{m}s_{0}\tau_{z}c_{m}$ into the Hamiltonian $H$. $W_{1}$ depict the disorder strength and $\omega_{m}$ is uniformly distributed within $[-0.5, 0.5]$. In Fig.~\ref{fig3}(a), with $g$ fixed at $0.5$, we plot the energy of the Hamiltonian $H+H_{W1}$ as a function of the disorder strength $W_{1}$. To improve computational efficiency, we employed a sparse-matrix method to solve the Hamiltonian, which limits the resolution to energies close to zero. Under the influence of the disorder $H_{W1}$, the system preserves both particle-hole symmetry $P$ and the composite symmetry $Sm_{z}$. As a result, the zero-energy states remain robust when the disorder strength $W_{1}$ is weak. Beyond a critical disorder strength $W_{1}\approx1.8$, the gap closes. As shown in Fig.~\ref{fig3}(b), the quadrupole moment $Q_{xy}$ maintains its quantized value of $0.5$ for weak disorder strength $W_{1}$. The spectrum at disorder strength $W_{1}=0.2$ is displayed in Fig.~\ref{fig3}(c). Although the fourfold degeneracy of bulk states is lifted, the zero-energy modes inside the gap remain robust. Furthermore, as shown in Fig.~\ref{fig3}(d), these zero-energy states are localized in a zero-dimensional form on both the inner and outer boundaries.

\subsection{Tunable number of zero-energy cornerlike states}
\label{SecIIC}
\begin{figure}[t]
	\includegraphics[width=0.48\textwidth]{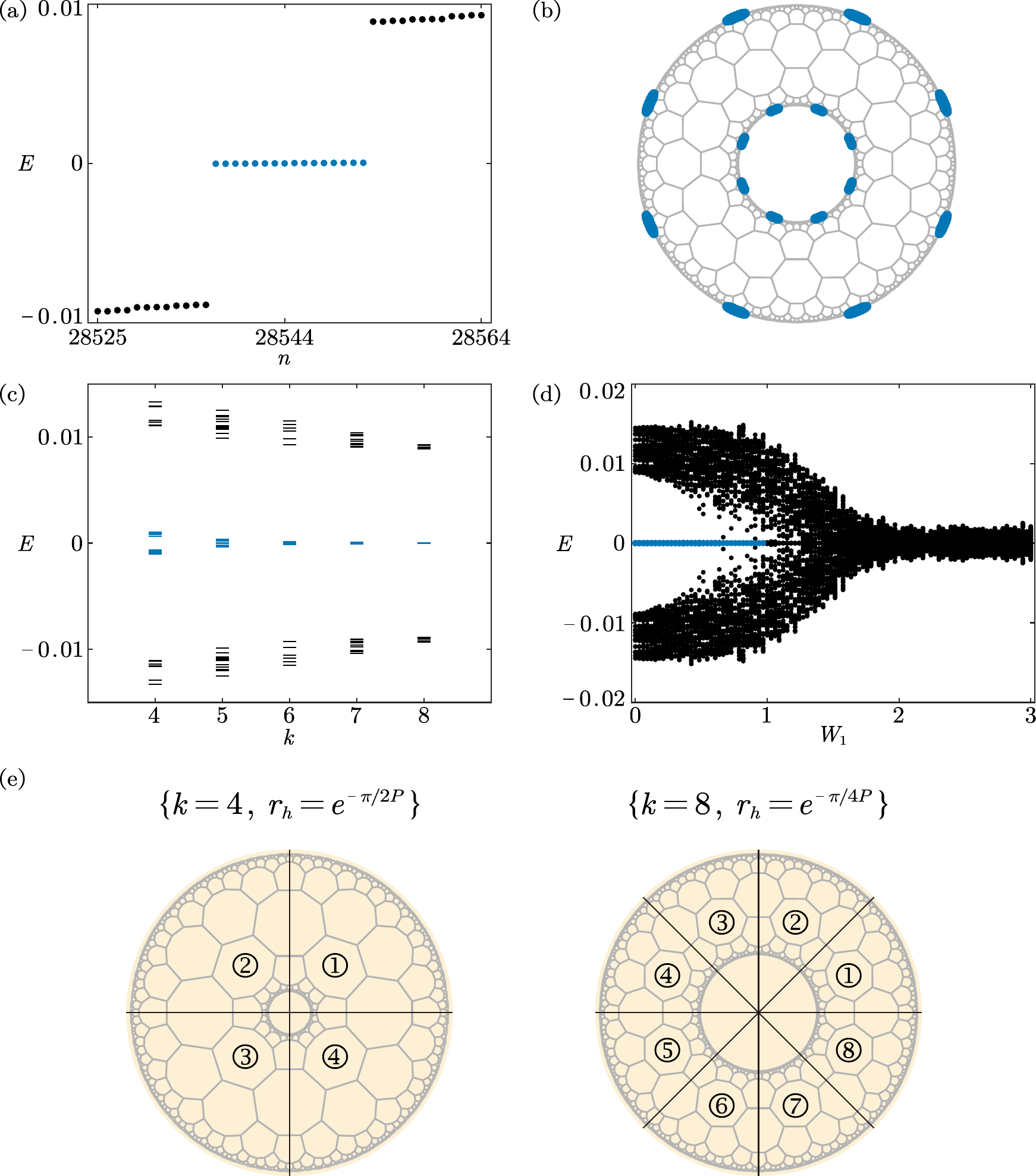} \caption{(a) Energy spectrum of the Hamiltonian $H$ in the $\{8, 3, 8\}$ lattice when $g=0.5$ and $\eta=4$. (b) The probability distribution of the sixteen zero-energy eigenstates marked with blue dots in (a). (c) Energy of the Hamiltonian $H$ for different $k$ when $g=0.5$ and $\eta=4$. (d) Energy of the Hamiltonian $H+H_{W1}$ as a function of the disorder strength $W_{1}$ when $k=8$. (e) Comparison of structures between the $\{8,3,4\}$ and $\{8,3,8\}$ lattices. The yellow sectors mark the repeating units. Here, we take the parameters $M=-1$, $t_{1}=t_{2}=1$, $g=0.5$, and $\eta=4$.}%
	\label{fig4}
\end{figure}

In type-I hyperbolic lattices, the number of zero-energy localized states can be tuned by varying the variation period of the Wilson mass term~\cite{PhysRevB.107.125302, PhysRevB.107.184201}. To explore how the variation period of the Wilson mass affects the zero-energy cornerlike states in type-II hyperbolic lattices, we adjust the parameter $\eta$ in the Wilson mass term to $\eta=4$. Diagonalizing the Hamiltonian $H$ yields the spectrum shown in Fig.~\ref{fig4}(a). One can see that the number of zero-energy cornerlike states increases to sixteen, i.e., $4\eta$ modes. Figure~\ref{fig4}(b) shows that these zero-energy states are localized on both the inner and outer boundaries. Separately, we examine the effect of disorder on systems hosting sixteen zero-energy cornerlike states. In Fig.~\ref{fig4}(d), we show the energy of the Hamiltonian $H+H_{W1}$ for the $\{8, 3, 8\}$ lattice versus the disorder strength $W_{1}$. The results demonstrate that, under weak disorder, the zero-energy cornerlike states remain stable inside the boundary-state gap. The gap closes once the disorder exceeds a critical strength $W_{1}\approx1$.

Furthermore, we examine how the zero-energy cornerlike states evolve in type-II hyperbolic lattices with different values of $k$. Figure~\ref{fig4}(c) shows the near-zero-energy states as a function of $k$ for $g=0.5$ and $\eta=4$. Owing to finite-size effects, the cornerlike states overlap and hybridize at small $k$, opening a small gap. As $k$ increases, the inner-circle radius of the type-II hyperbolic lattice $r_{\rm{in}}=r_{h}^{2}=e^{-4\pi/kP}$ enlarges in Fig.~\ref{fig4}(e), the cornerlike states recover zero energy. Due to the fact that the numbers of sites on the inner and outer boundaries of type-II hyperbolic lattices are equal, when $k$ is small, the cornerlike states on the boundaries overlap with each other, causing those localized on the inner and outer boundaries to simultaneously deviate from zero energy as shown in Fig.~\ref{fig4}(c). As the value of $k$ increases, corresponding to an increase in the number of sites on both boundaries, the cornerlike states localized on the inner and outer boundaries gradually and simultaneously recover zero energy. This characteristic of type-II hyperbolic lattices contrasts with that of type-I hyperbolic lattices with an introduced central hole and square lattices confined to an annular geometry, where the number of sites on the outer boundary substantially exceeds that on the inner boundary. Consequently, the cornerlike states localized on each boundary respond differently to changes in system size. For instance, the outer boundary may host a sufficient number of sites so that its cornerlike states reside exactly at zero energy, whereas the inner boundary, having far fewer sites, hosts cornerlike states that deviate from zero energy (see Appendix~\ref{AppendixC} for more details).

We note that the real-space quadrupole moment $Q_{xy}$ in Sec.~\ref{SecIIB} characterizes precisely the topological phase with eight zero-energy cornerlike states.  When calculating the quadrupole moment for the sample with a single boundary, we place the sample center at the origin of the coordinate axes. If an odd number of zero-energy corner states resides in any quadrant, the quadrupole moment of the system is $Q_{xy}=0.5$. If an even number is found in any quadrant, the quadrupole moment is zero~\cite{PhysRevB.110.L121301}. In type-II hyperbolic lattices, one can verify the quadrupole moment by counting the zero-energy cornerlike states in any quadrant of either the inner or the outer boundary. For the type-II hyperbolic lattice with $\eta=2$, each quadrant of the outer (inner) boundary contains exactly one zero-energy cornerlike state, yielding a quadrupole moment of $Q_{xy}=0.5$. For the type-II hyperbolic lattice with $\eta=4$, which hosts sixteen zero-energy cornerlike states, each quadrant of the outer boundary contains two zero-energy cornerlike states, resulting in a vanishing quadrupole moment. Recent studies have suggested that a coordinate transformation can be applied so that an odd number of corner states falls within each quadrant in the HOTI with eight zero-energy corner states, thereby enabling the computation of a non-zero generalized quadrupole moment $Q_{x^{\prime}y^{\prime}}$ using the transformed coordinates~\cite{10.21468/SciPostPhys.15.5.193, PhysRevB.109.195301}.

\begin{figure}[t]
	\includegraphics[width=0.48\textwidth]{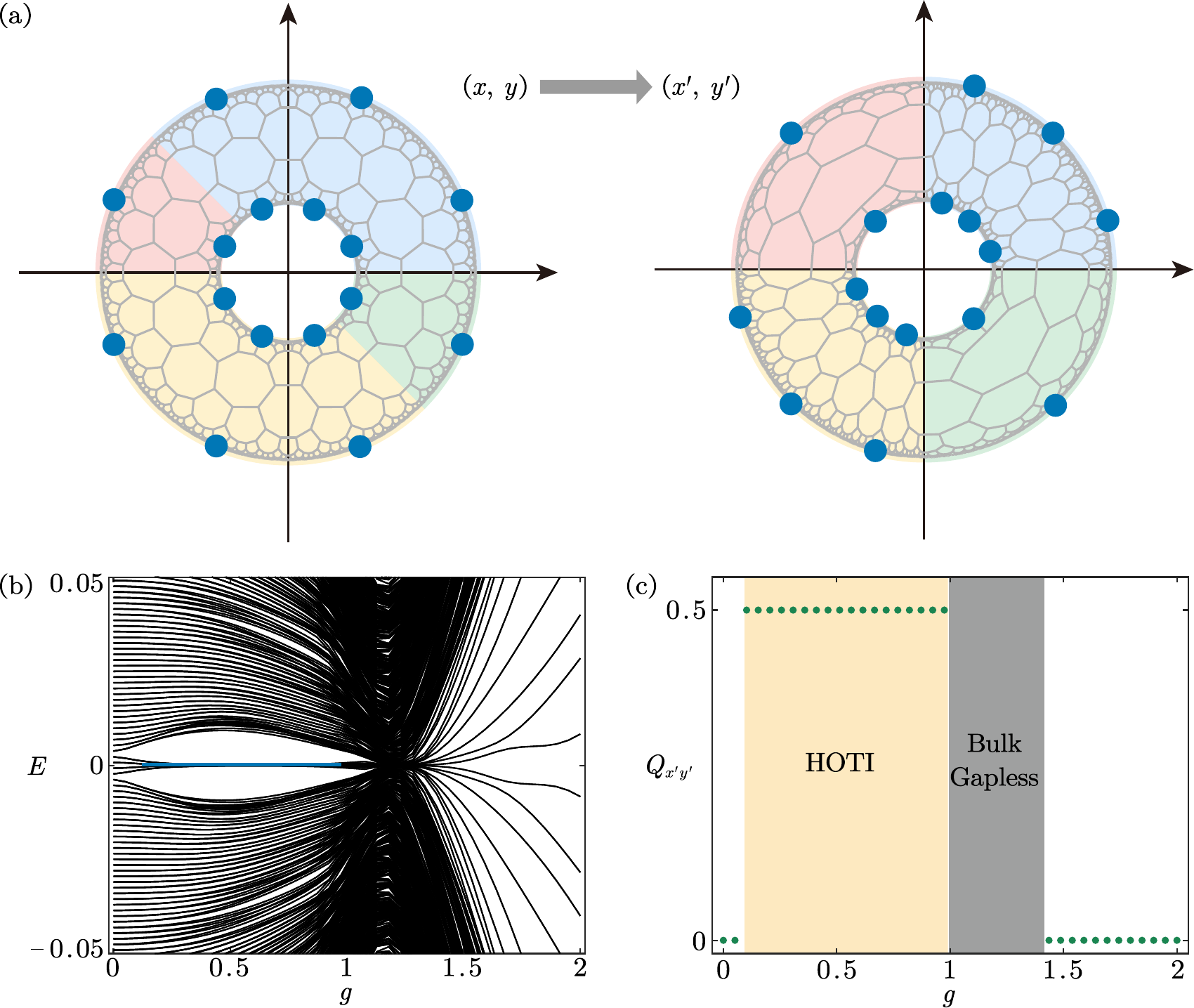} \caption{(a) Schematics of how to change site positions in the $\{8, 3, 8\}$ lattice. Left panel: A type-II hyperbolic lattice with sixteen zero-energy cornerlike states. The coordinate $(x_{n}, y_{n})$ of the $n$th site is expressed on the complex plane as $r_{n} e^{i\theta_{n}}$ with $-\pi \le \theta_{n} < \pi$. To correctly compute the generalized quadrupole moment $Q_{x^{\prime}y^{\prime}}$, we apply a coordinate transformation defined piecewise: for $0 \le \theta_n < 3\pi/4$, $\theta_{n}^{\prime} = \frac{2}{3}\theta_n$; for $3\pi/4 \le \theta_n < \pi$, $\theta_{n}^{\prime} = 2\theta_n - \pi$; for $-\pi \le \theta_n < -\pi/4$, $\theta_{n}^{\prime} = \frac{2}{3}\theta_n - \pi/3$; for $-\pi/4 \le \theta_n < 0$, $\theta_{n}^{\prime} = 2\theta_n$. The transformed coordinate is $r_n e^{i\theta_n^{\prime}}$ or $(x^{\prime}, y^{\prime})$. Right panel: The type-II hyperbolic lattice after the coordinate transformation. (b) Energy of the Hamiltonian $H$ as a function of the Wilson mass $g$ in the $\{8, 3, 6\}$ lattice. (c) The generalized quadrupole moment $Q_{x^{\prime}y^{\prime}}$ as a function of the Wilson mass $g$ in the $\{8, 3, 6\}$ lattice. Here, we take the parameters $M=-1$, $t_{1}=t_{2}=1$, and $\eta=4$.}%
	\label{fig5}
\end{figure}

As shown in the left panel of Fig.~\ref{fig5}(a), each quadrant of the outer (inner) boundary contains two zero-energy cornerlike states, resulting in a vanishing quadrupole moment. To characterize the topology of the system when $\eta=4$, we introduce a generalized quadrupole moment $Q_{x^{\prime}y^{\prime}}$. In the left panel, the type-II hyperbolic lattice is partitioned into four sectors, marked blue, red, yellow, and green. When expressed in the complex plane, the coordinate of site $n$ is written as $r_{n} e^{i\theta_{n}}$ or $(x_{n}, y_{n})$. We perform a coordinate transformation that compresses the blue sector into the first quadrant, expands the red sector to fill the second quadrant, compresses the yellow sector into the third quadrant, and expands the green sector to fill the fourth quadrant, as illustrated in the right panel of Fig.~\ref{fig5}(a). After this transformation, every quadrant contains an odd number of zero-energy cornerlike states. The transformation is defined piecewise: when the site $n$ locates in the blue sector, i.e., $0 \le \theta_n < 3\pi/4$, then $\theta_{n}^{\prime} = \frac{2}{3}\theta_n$; when the site $n$ locates in the red sector, i.e., $3\pi/4 \le \theta_n < \pi$, then $\theta_{n}^{\prime} = 2\theta_n - \pi$; when the site $n$ locates in the yellow sector, i.e., $-\pi \le \theta_n < -\pi/4$, then $\theta_{n}^{\prime} = \frac{2}{3}\theta_n - \pi/3$; when the site $n$ locates in the green sector, i.e., $-\pi/4 \le \theta_n < 0$, then $\theta_{n}^{\prime} = 2\theta_n$. The transformed coordinate is $r_n e^{i\theta_n^{\prime}}$ or $(x_{n}^{\prime}, y_{n}^{\prime})$. Using the transformed coordinates $(x_{n}^{\prime}, y_{n}^{\prime})$, the generalized quadrupole moment $Q_{x^{\prime}y^{\prime}}$ is computed via Eq.~(\ref{eqQxy1}).

Numerical calculation shows that the type-II hyperbolic lattice hosting sixteen zero-energy cornerlike states yields a nonzero generalized quadrupole moment $Q_{x^{\prime}y^{\prime}}=0.5$. In Figs.~\ref{fig5}(b) and \ref{fig5}(c), we show the energy spectrum and the generalized quadrupole moment $Q_{x^{\prime}y^{\prime}}$ as functions of the Wilson mass $g$ when $\eta=4$. Upon introducing a finite $g$, the boundary-state gap is opened. Within this gap, zero-energy cornerlike states emerge, signaling a phase transition from a quantum spin Hall insulator to a HOTI characterized by a nonzero generalized quadrupole moment $Q_{x^{\prime}y^{\prime}}=0.5$. When the Wilson mass increases to approximately $g=1$, the bulk gap closes. In the bulk gapless regime, as indicated by the grey region in Fig.~\ref{fig5}(c), the generalized quadrupole moment is not well-defined. When $g>1.4$, the system enters a trivial insulator phase. In addition to type-II hyperbolic lattices with $\eta = 4$, we find that the generalized quadrupole moment $Q_{x^{\prime}y^{\prime}}$ can also be employed to characterize those with $\eta = 6$.

\subsection{Spatial control of zero-energy cornerlike states}
\label{SecIID}
We note that the zero-energy cornerlike states are pinned to fixed boundary positions and ask whether this localization can be controlled. To that end, we modify the Hamiltonian $H$ in the following way:
\begin{align}
H_{\rm{R}}=&-\frac{1}{2}\sum_{\left <m,n \right >}c_{m}^{\dagger}it_{1}\left [s_{z}\tau_{x}\cos(\phi_{mn})+s_{0}\tau_{y}\sin(\phi_{mn})\right ]c_{n}\nonumber \\
&-\frac{1}{2}\sum_{\left <m,n \right >}c_{m}^{\dagger}t_{2}s_{0}\tau_{z}c_{n}+\sum_{m}(M+2t_{2})c_{m}^{\dagger}s_{0}\tau_{z}c_{m}\nonumber \\
&+\frac{g}{2}\sum_{\left <m,n \right >}c_{m}^{\dagger}\cos(\eta\phi_{mn})s_{x}\tau_{x}c_{n},
\label{eqR}
\end{align}
where $\phi_{mn}$ is defined in relation to the spatial positions of sites $m$ and $n$. Specifically, when both sites reside on the inner-ring of the Poincar\'{e} ring, $\phi_{mn}=\theta_{mn}+\phi_{\rm{in}}$. When both are on the outer-ring, $\phi_{mn}=\theta_{mn}+\phi_{\rm{out}}$.

\begin{figure}[t]
	\includegraphics[width=0.48\textwidth]{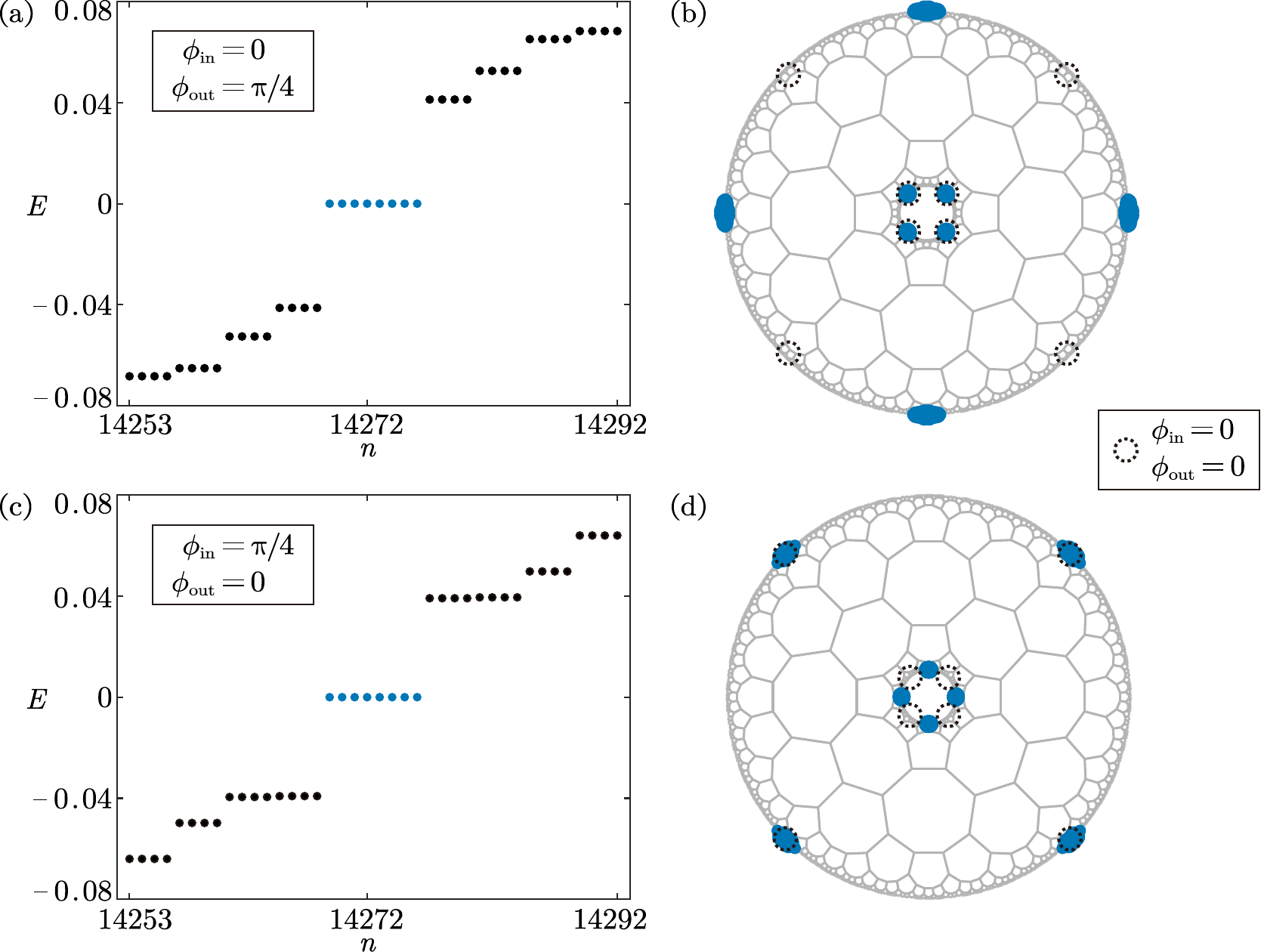} \caption{(a) Energy spectrum of the Hamiltonian $H_{\rm{R}}$ in the $\{8, 3, 4\}$ lattice when $\phi_{\rm{in}}=0$ and $\phi_{\rm{out}}=\pi/4$. (b) The probability distribution of the eight zero-energy eigenstates marked with blue dots in (a). (c) Energy spectrum of the Hamiltonian $H_{\rm{R}}$ in the $\{8, 3, 4\}$ lattice when $\phi_{\rm{in}}=\pi/4$ and $\phi_{\rm{out}}=0$. (d) The probability distribution of the eight zero-energy eigenstates marked with blue dots in (c). In (b) and (d), the dashed circles mark the localized positions of the zero-energy cornerlike states in the system with $\phi_{\rm{in}}=0$ and $\phi_{\rm{out}}=0$. Here, we take the parameters $M=-1$, $t_{1}=t_{2}=1$, $g=0.5$, and $\eta=2$.}%
	\label{fig6}
\end{figure}

For the system with $\phi_{\rm{in}}=0$ and $\phi_{\rm{out}}=0$, the results reproduce those in Sec.~\ref{SecIIA}. When $\phi_{\rm{in}}=0$ and $\phi_{\rm{out}}=\pi/4$, eight zero-energy states appear inside the boundary-state gap, as shown in Fig.~\ref{fig6}(a). Their probability distributions [Fig.~\ref{fig6}(b)] reveal that the cornerlike states in the inner-ring remain at the same positions as in Fig.~\ref{fig2}(d), whereas the cornerlike states in the outer-ring are rotated by $\pi/4$. In Fig.~\ref{fig6}(b), the dashed circles mark the localized positions of the zero-energy cornerlike states for the system with $\phi_{\rm{in}}=0$ and $\phi_{\rm{out}}=0$. Similarly, the energy spectrum and the probability distribution of the zero-energy states for the system with $\phi_{\rm{in}}=\pi/4$ and $\phi_{\rm{out}}=0$ are shown in Figs.~\ref{fig6}(c) and \ref{fig6}(d), respectively. We find that the cornerlike states in the inner-ring rotate by $\pi/4$ while the cornerlike states in the outer-ring stay at the same positions as in Fig.~\ref{fig2}(d). In both cases the conventional quadrupole moment $Q_{xy}$ vanishes.

To characterize the topology of these rotated patterns, we introduce a generalized quadrupole moment based on a coordinate transformation~\cite{10.21468/SciPostPhys.15.5.193, PhysRevB.110.L121301}. Specifically, for the system with $\phi_{\rm{in}}=0$ and $\phi_{\rm{out}}=\pi/4$, we rotate the outer-ring coordinates counter-clockwise by $\pi/4$ while keeping the inner-ring coordinates fixed. For the system with $\phi_{\rm{in}}=\pi/4$ and $\phi_{\rm{out}}=0$, we rotate the inner-ring coordinates by $\pi/4$ and leave the outer-ring coordinates unchanged. Using the transformed coordinates $(x^{r}, y^{r})$ in Eq.~(\ref{eqQxy1}), the generalized quadrupole moment exhibits the quantized value $Q_{x^{r}y^{r}} = 0.5$.

\vspace{-0.1cm}
\section{Benalcazar-Bernevig-Hughes Model in type-II hyperbolic lattices}
\label{SecIII}
\begin{figure}[t]
	\includegraphics[width=0.48\textwidth]{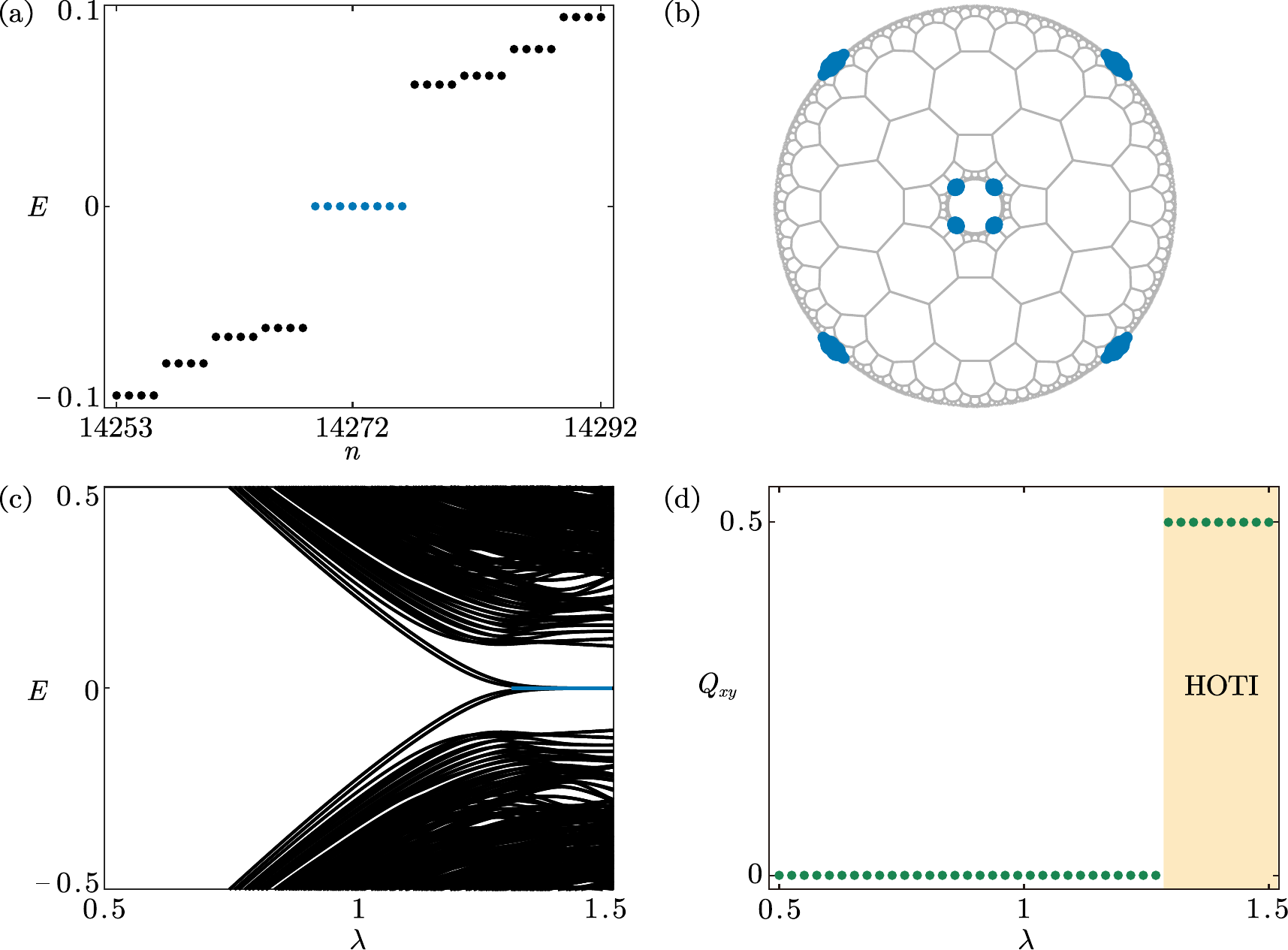} \caption{(a) Energy spectrum of the Hamiltonian $H_{\rm{BBH}}$ in the $\{8, 3, 4\}$ lattice when $\lambda=1.5$. (b) The probability distribution of the eight zero-energy eigenstates marked with blue dots in (a). (c) Energy of the Hamiltonian $H_{\rm{BBH}}$ as a function of $\lambda$. (d) The quadrupole moment $Q_{xy}$ as a function of $\lambda$.  Here, we take the parameters $\gamma=1$ and $k=4$.}%
	\label{fig7}
\end{figure}

In this section, we investigate the phase transition of the BBH model in type-II hyperbolic lattices. Here, we apply the BBH model to the type-II hyperbolic lattice. The Hamiltonian can be described by the following expression \cite{doi:10.1126/science.aah6442, PhysRevLett.124.036803}:
\begin{align}
H_{\rm{BBH}}&=\gamma\sum_{m}c_{m}^{\dagger}(\Gamma_{2}+\Gamma_{4})c_{m}\nonumber\\
&+\frac{\lambda}{2}\sum_{\left <m,n \right >}c_{m}^{\dagger}[ |\cos(\theta_{mn})|\Gamma_{4}-i \cos(\theta_{mn})\Gamma_{3}\nonumber\\
&+|\sin(\theta_{mn})|\Gamma_{2}-i \sin(\theta_{mn})\Gamma_{1}] c_{n},
\label{BBH}
\end{align}
where $c_{m}^{\dagger}$ and $c_{m}$ are the creation and annihilation operators of electrons on site $m$. $\theta_{mn}$ represents the polar angle of the vector from the site $n$ to the site $m$ in the Poincar\'{e} ring. Gamma matrices are given by $\Gamma_{1}=-\sigma_{y}\sigma_{x}$, $\Gamma_{2}=-\sigma_{y}\sigma_{y}$, $\Gamma_{3}=-\sigma_{y}\sigma_{z}$, and $\Gamma_{4}=\sigma_{x}\sigma_{0}$. $\sigma_{x,y,z}$ are the Pauli matrices acting on the sublattice, $\sigma_{0}$ is the identity matrix. $\gamma$ represents the hopping amplitude between the sublattices of the same site. In subsequent calculations, we set $\gamma=1$. $\lambda$ represents the hopping amplitude between the nearest-neighbor sites. The Hamiltonian $H_{\rm BBH}$ respects the time-reversal symmetry $\mathcal{T}=\mathcal{K}$, the particle-hole symmetry $\mathcal{P}=\sigma_{z}\sigma_{0}\mathcal{K}$, and the chiral symmetry $\mathcal{S}=\mathcal{PT}=\sigma_{z}\sigma_{0}$.

When $\lambda=1.5$, the energy spectrum of the Hamiltonian is shown in Fig.~\ref{fig7}(a). We observe a gap around zero energy, inside which there exist eight degenerate zero-energy states. Similar to the HOTI discussed in the previous section for type-II hyperbolic lattices, these eight zero-energy states are localized in a zero-dimensional form on both boundaries of the Poincar\'{e} ring, as illustrated in Fig.~\ref{fig7}(b). To investigate the topological phase transition of the BBH model on a type-II hyperbolic lattice, we compute the energy of the Hamiltonian as a function of $\lambda$ as shown in Fig.~\ref{fig7}(c). It can be seen that the bulk states remain gapped for small $\lambda$. When $\lambda$ exceeds a critical value $1.3$, zero-energy states emerge inside the bulk gap, signaling a transition from a trivial insulator to a HOTI hosting zero-energy cornerlike states. Similar to Sec.~\ref{SecII}, we employ the real-space quadrupole moment $Q_{xy}$ to characterize the topological properties of the system. In Fig.~\ref{fig7}(d), we show the variation of the quadrupole moment $Q_{xy}$ with the system parameter $\lambda$. Comparing with Fig.~\ref{fig7}(c), one observes that $Q_{xy}=0.5$ corresponds to the HOTI phase hosting zero-energy cornerlike states, whereas $Q_{xy}=0$ indicates a trivial insulator phase.

\begin{figure}[t]
	\includegraphics[width=0.48\textwidth]{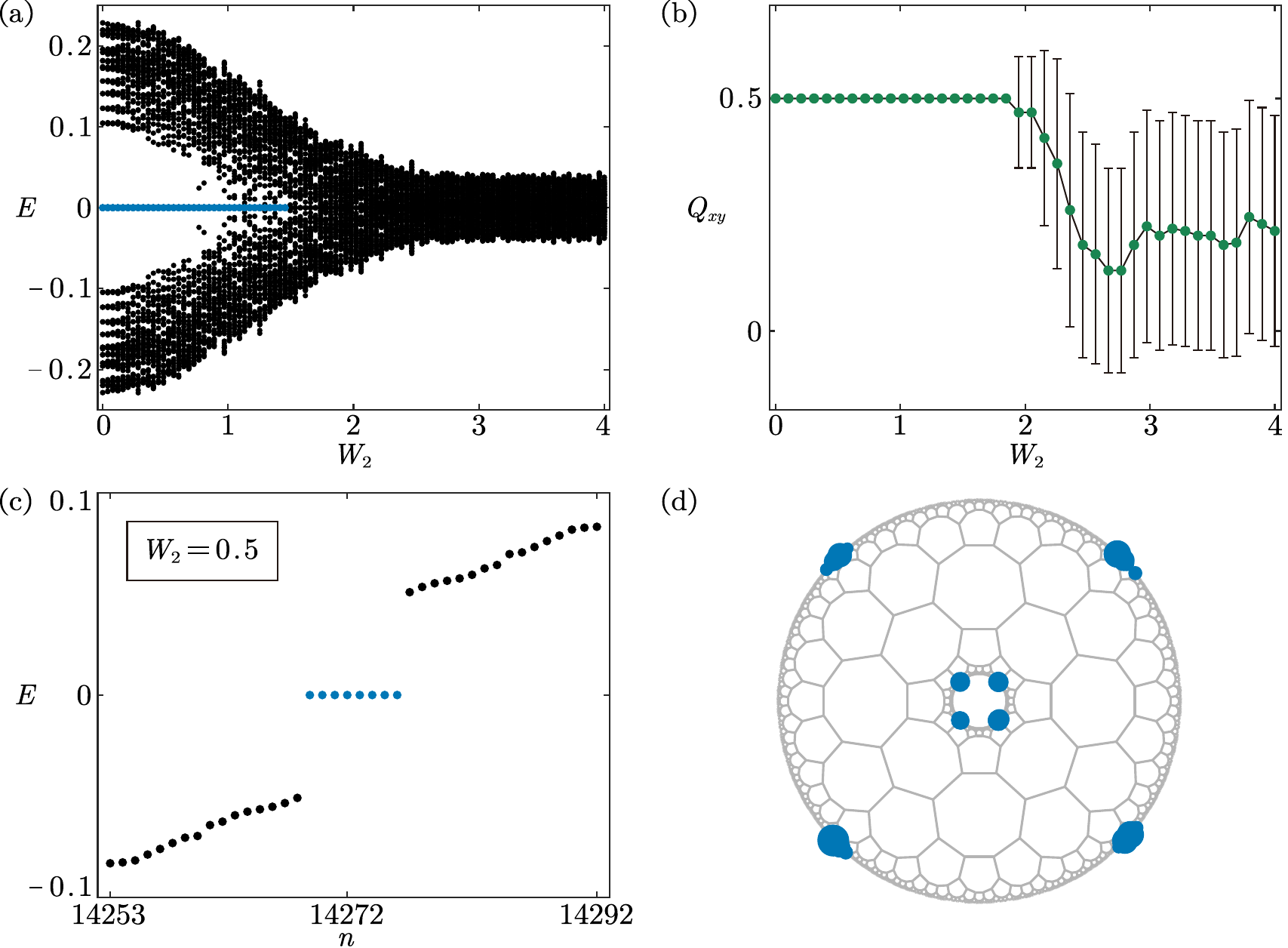} \caption{(a) Energy of the Hamiltonian $H_{\rm{BBH}}+H_{W2}$ as a function of the disorder strength $W_{2}$. (b) The quadrupole moment $Q_{xy}$ as a function of the disorder strength $W_{2}$. The error bar represents the standard deviation of 100 samples. (c) Energy spectrum of the Hamiltonian $H_{\rm{BBH}}+H_{W2}$ when $W_{2}=0.5$. (d) The probability distribution of the eight zero-energy eigenstates marked with blue dots in (c). Here, we take the parameters $\gamma=1$, $\lambda=1.5$, and $k=4$.}%
	\label{fig8}
\end{figure}

To further verify the robustness of the zero-energy localized states against disorder, we introduce the disorder term $H_{W2}=W_{2}\sum_{m}c_{m}^{\dagger}\omega_{m}(\Gamma_{2}+\Gamma_{4})c_{m}$ into the BBH model $H_{\rm{BBH}}$. Figures~\ref{fig8}(a) and \ref{fig8}(b) show the evolution of the energy $E$ and the quadrupole moment $Q_{xy}$ with respect to the disorder strength $W_2$, respectively. The results confirm that the zero-energy states remain robust against weak disorder, and the quadrupole moment retains its quantized value $Q_{xy}=0.5$ in this regime. In addition, we present the energy spectrum at disorder strength $W_{2}=0.5$ together with the probability distribution of the zero-energy states, as shown in Figs.~\ref{fig8}(c) and \ref{fig8}(d). Under weak disorder, the degeneracy of the bulk states is lifted, while the zero-energy states remain robust. It is worth noting that when the BBH model is applied to type-II hyperbolic lattices, the number of zero-energy cornerlike states is fixed at eight, while their spatial positions can be tuned. Analogous to the discussion in Sec.~\ref{SecIID}, by replacing the polar angle $\theta_{mn}$ in Eq.~(\ref{BBH}) with $\phi_{mn}$, the spatial distribution of the cornerlike states on both boundaries can be controlled through $\phi_{\rm{in}}$ and $\phi_{\rm{out}}$, respectively.

\section{Conclusion}
\label{Conclusion}
In this work, we reveal the HOTI phases in type-II hyperbolic lattices based on the modified BHZ model and the BBH model. In contrast to type-I lattices, type-II hyperbolic lattices feature zero-energy cornerlike states on both boundaries whose locations can be independently controlled. Furthermore, owing to the identical number of sites on the inner and outer boundaries of type-II hyperbolic lattices, the cornerlike states localized on both boundaries exhibit a uniform response to variations in system size. This behavior contrasts with that of type-I hyperbolic lattices with an artificially introduced central hole or Euclidean lattices confined to an annular geometry.

First, in the modified BHZ model, the number of zero-energy cornerlike states is controlled by the variation period of the Wilson mass term. For $\eta=2$, the system hosts eight zero-energy cornerlike states, with four localized on the inner and outer boundaries. The HOTI phase with eight zero-energy cornerlike states is characterized by the quadrupole moment. We further show that the spatial positions of the zero-energy cornerlike states can be controlled, enabling their relative shift between the inner and outer boundaries. The topology of such configurations is captured by a generalized quadrupole moment. When $\eta$ is increased to 4, the number of zero-energy cornerlike states increases to sixteen, corresponding to eight zero-energy cornerlike states on the inner and outer boundaries. It is noted that the HOTI phase with sixteen zero-energy cornerlike states is characterized by the generalized quadrupole moment. Moreover, we find that increasing the structural parameter $k$ can mitigate finite-size effects. Furthermore, we demonstrate that the zero-energy cornerlike states are topologically robust against weak disorder.

Second, in the BBH model, the system can transition from a trivial insulator phase to the HOTI phase with $Q_{xy}=0.5$ by tuning the system parameter $\lambda$. In the nontrivial type-II hyperbolic lattice, the system hosts eight zero-energy cornerlike states, and each boundary supports four zero-energy cornerlike states. In the presence of weak disorder, these zero-energy cornerlike states remain stable.

Type-I Hyperbolic lattices with negative curvature can be realized in the Poincar\'{e} disk embedded in the Euclidean plane. Recent experimental studies have reported that circuit quantum electrodynamics~\cite{Kollar2019}, electronic circuits~\cite{Lenggenhager2022, Zhang2022, 10.1038/s41467-023-36359-6, PhysRevB.107.165145}, and photonic systems~\cite{10.1021/acsphotonics.4c01184, 10.1038/s41467-024-46035-y, PhysRevB.110.155123} provide viable platforms for implementing type-I hyperbolic lattices. Although type-II hyperbolic lattices are constructed on the Poincar\'{e} ring, they do not introduce additional technical challenges in practice. We expect that higher-order topological states in type-II hyperbolic lattices can be realized in these classical settings.

\section*{Acknowledgments}
We acknowledge the support of the NSFC (Grants No.~U25D8012, No.~12074107, No.~12304195, No.~12304539, No.~12504232, No.~12504560), the Chutian Scholars Program in Hubei Province, the Hubei Provincial Natural Science Foundation (Grants No.~2025AFA081, No.~2022CFA012, No.~2025AFB397), the Wuhan city key R\&D program (Grant No. 2025050602030069), the program of outstanding young and middle-aged scientific and technological innovation team of colleges and universities in Hubei Province (Grant No. T2020001), the key project of Hubei provincial department of education (Grant No. D20241004), the Guangdong Provincial Quantum Science Strategic Initiative (GDZX2401001), the original seed program of Hubei university, the Research Foundation of Hubei Educational Committee (Grant No. Q20231807), and the Postdoctor Project of Hubei Province (Grant No. 2025HBBSHCXB015).

\appendix

\section{The influence of radial expansion on the phase transition point}
\label{AppendixA}
System expansion in type-II hyperbolic lattices can occur by increasing the structural parameter $k$ or by radial expansion (increasing the number of layers between boundaries). We use the base layer number $L$ to quantify the radial expansion of type-II hyperbolic lattices. The base layer number $L$ refers to the number of layers in the type-I hyperbolic lattice. The type-I hyperbolic lattice can be transformed into a type-II hyperbolic lattice via a conformal projection~\cite{10.1038/s42005-025-01990-w}. As the base layer number $L$ increases, the number of layers between the inner and outer boundaries in the type-II lattice correspondingly increases.

\begin{figure}[h]
\centering
	\includegraphics[width=0.48\textwidth]{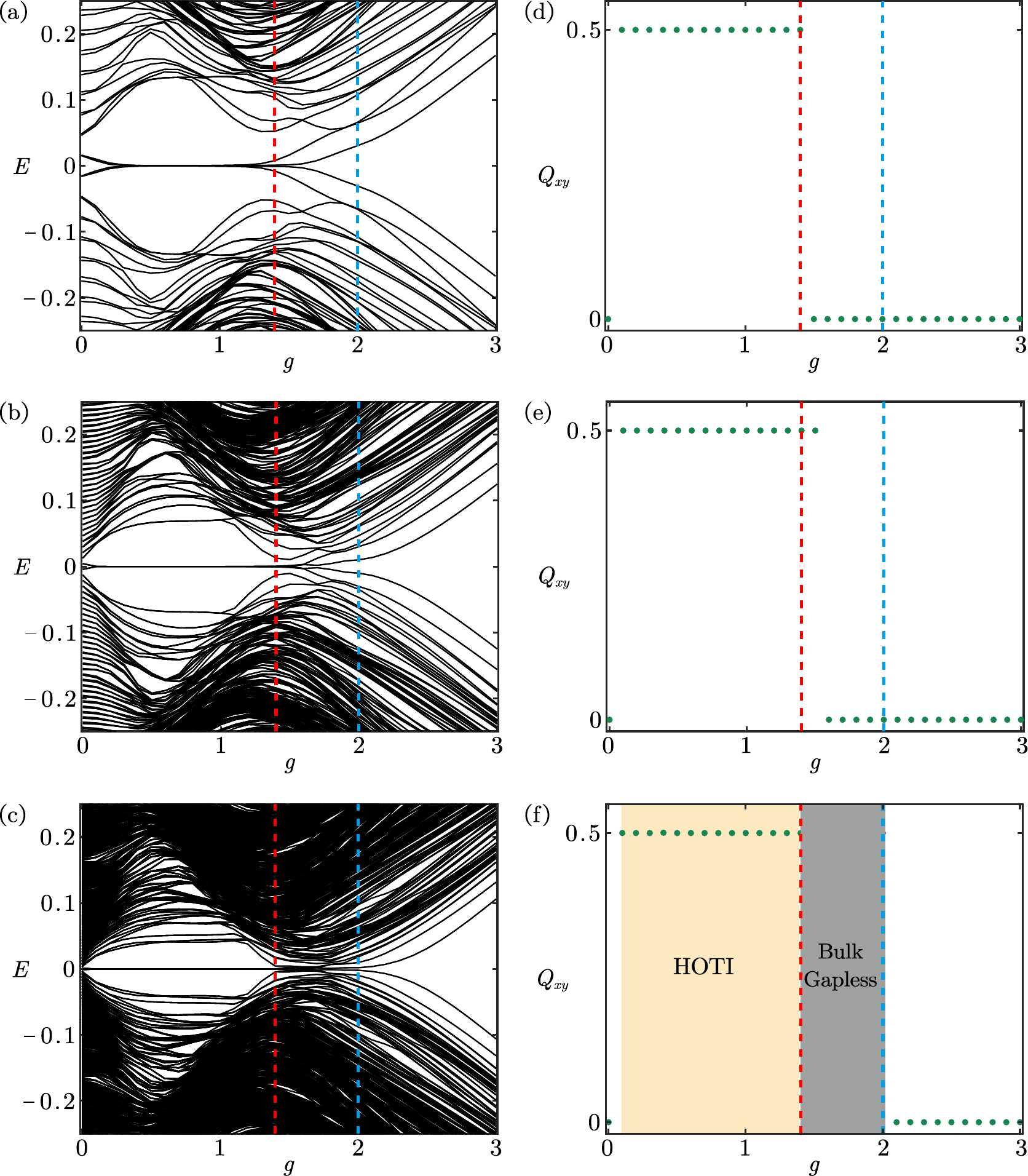}
\centering
\caption{(a), (b), and (c) show the energy spectrum of the Hamiltonian $H$ as a function of the Wilson mass $g$ in the $\{p = 8, q = 3, k = 4\}$ lattice with base layer numbers $L = 3$, $L = 4$, and $L = 5$, respectively. (d), (e), and (f) show the quadrupole moment $Q_{xy}$ of the Hamiltonian $H$ as a function of the Wilson mass $g$ in the $\{p = 8, q = 3, k = 4\}$ lattice with base layer numbers $L = 3$, $L = 4$, and $L = 5$, respectively. The red and cyan dashed lines indicate $g = 1.4$ and $g = 2$, respectively. Here, we take $M = -1$ and $\eta = 2$.}%
	\label{figA}
\end{figure}

In Figs.~\ref{figA}(a)-\ref{figA}(c), we present the energy spectrum of the $\{8, 3, 4\}$ lattice as a function of the Wilson mass $g$ for base layer numbers $L = 3$, $L = 4$, and $L = 5$, respectively. The red and cyan dashed lines indicate $g = 1.4$ and $g = 2$, respectively. It is evident that for $0 < g < 1.4$, the system resides in the HOTI phase with the nonzero quadrupole moment $Q_{xy}=0.5$ [as shown in Figs.~\ref{figA}(d)-\ref{figA}(f)]. As $L$ increases, the gap associated with the zero-energy cornerlike states gradually narrows. However, it remains unclear whether this gap vanishes in the thermodynamic limit or converges to a finite nonzero value. In the interval $1.4 < g < 2$, the bulk gap progressively closes with increasing $L$, eventually leading to a transition into a metallic phase.

\section{The influence of curvature on the phase transition point}
\label{AppendixB}
\begin{figure}[t]
\centering
	\includegraphics[width=0.48\textwidth]{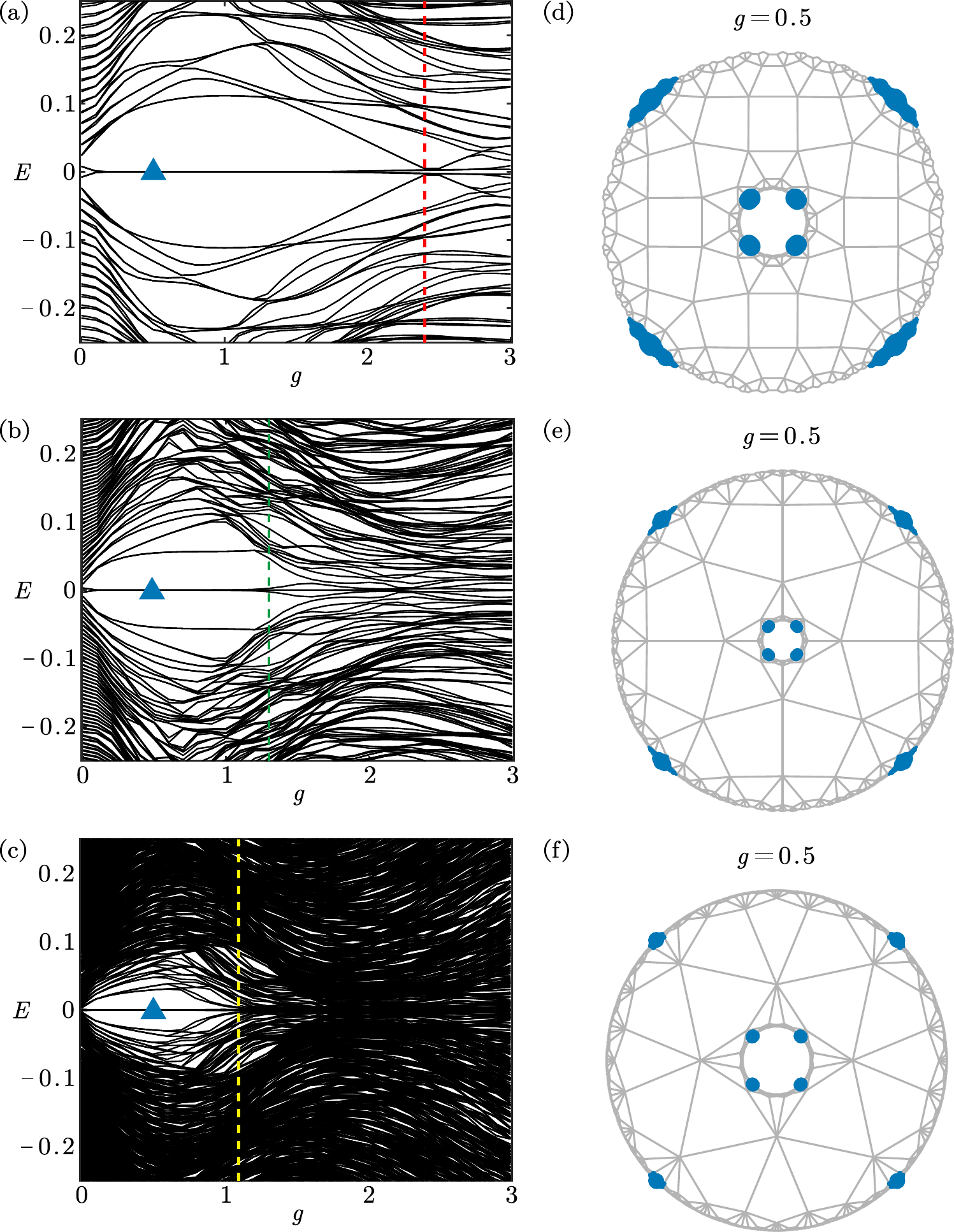}
\centering
\caption{(a) Energy of the Hamiltonian $H$ as a function of the Wilson mass $g$ in the $\{4, 5, 4\}$ lattice. The red dashed line indicates $g = 2.4$. (b) Energy of the Hamiltonian $H$ as a function of the Wilson mass $g$ in the $\{4, 6, 4\}$ lattice. The green dashed line indicates $g = 1.3$. (c) Energy of the Hamiltonian $H$ as a function of the Wilson mass $g$ in the $\{4, 8, 4\}$ lattice. The yellow dashed line indicates $g = 1.1$. (d), (e), and (f) show the probability distributions of the zero-energy states marked by blue triangles in (a), (b), and (c), respectively. Here, we take $M = -1$ and $\eta = 2$.}%
	\label{figB}
\end{figure}

In hyperbolic geometry, the coordination number $q$ directly dictates the constant negative curvature of the space. To verify whether type-II hyperbolic lattices based on other tilings also exhibit higher-order topological properties and to examine the influence of $q$ on the phase transition point, we construct type-II $\{p = 4, q\}$ lattices with different values of $q$.

Figures~\ref{figB}(a)-\ref{figB}(c) show the energy of the Hamiltonian $H$ as a function of the Wilson mass $g$ for the $\{p = 4, q = 5, k = 4\}$, $\{p = 4, q = 6, k = 4\}$, and $\{p = 4, q = 8, k = 4\}$ lattices, respectively. In these figures, the phase transition point from the HOTI to the metallic phase is marked by red, green, and yellow dashed lines, respectively. It can be observed that to the left of the phase transition point, the system hosts zero-energy states residing within the gap. Specifically, Figs.~\ref{figB}(d)-\ref{figB}(f) show the probability distributions of the zero-energy states marked by blue triangles in Figs.~\ref{figB}(a)-\ref{figB}(c), respectively, for $g = 0.5$. The topological nature of these zero-energy states is characterized by a quantized quadrupole moment $Q_{xy} = 0.5$, indicating that the system resides in the HOTI phase. As the value of $q$ increases, the phase transition point shifts, leading to a gradual shrinkage of the HOTI phase in the parameter space.

\section{The unique properties of higher-order topological states in type-II hyperbolic lattices}
\label{AppendixC}
\begin{figure}[h]
\centering
	\includegraphics[width=0.48\textwidth]{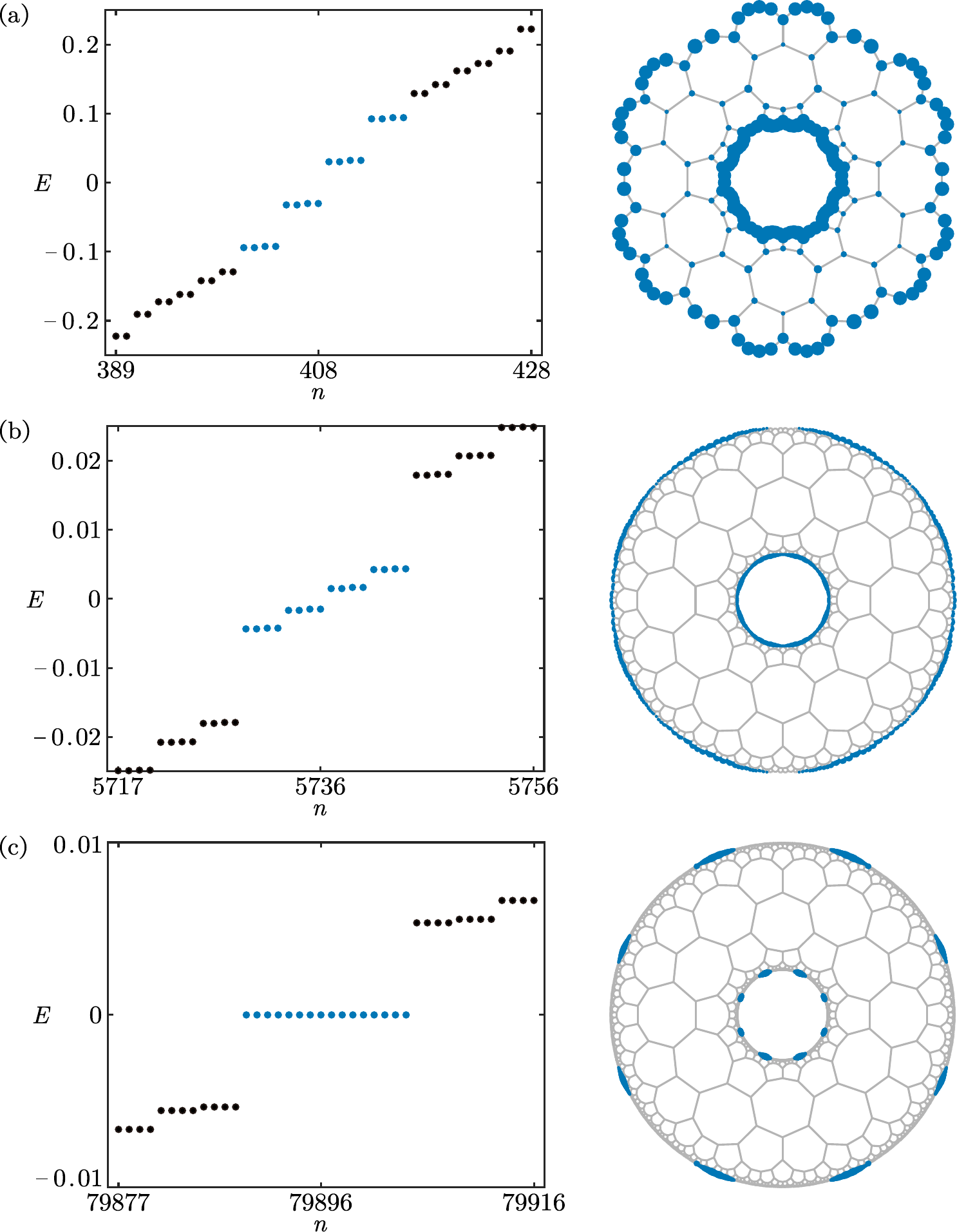}
\centering
\caption{The top panels of (a), (b), and (c) show the energy spectrum of the Hamiltonian $H$ applied to the $\{p = 8, q = 3, k = 6\}$ lattice with base layer numbers $L = 2$, $L = 4$, and $L = 6$, respectively. The bottom panels display the probability distributions of the eigenstates marked by blue dots in the top panels. Here, we take $M = -1$, $g = 0.5$, and $\eta = 4$.}%
	\label{figC1}
\end{figure}

In this section, we examine the differences between the cornerlike states in type-II hyperbolic lattices and those in type-I hyperbolic lattices or Euclidean square lattices.

First, we investigate the response of the corner states in type-II hyperbolic lattices to finite-size effects as the system size varies. For fixed $\{p = 8, q = 3, k = 6\}$, Fig.~\ref{figC1} displays the energy spectrum and the distributions of near-zero-energy states for type-II hyperbolic lattices with base layer numbers $L = 2$, $L = 4$, and $L = 6$. As the base layer number $L$ increases, the number of sites on both the inner and outer boundaries increases. When the base layer number $L$ is small, the near-zero-energy states localized on the inner and outer boundaries hybridize with each other, causing these eigenstates to deviate from zero energy as shown in Fig.~\ref{figC1}(a). As the base layer number $L$ increases, the cornerlike states localized on both the inner and outer boundaries gradually and simultaneously approach zero energy as shown in Figs.~\ref{figC1}(b) and \ref{figC1}(c).

\begin{figure*}[htbp]
\centering
	\includegraphics[width=0.9\textwidth]{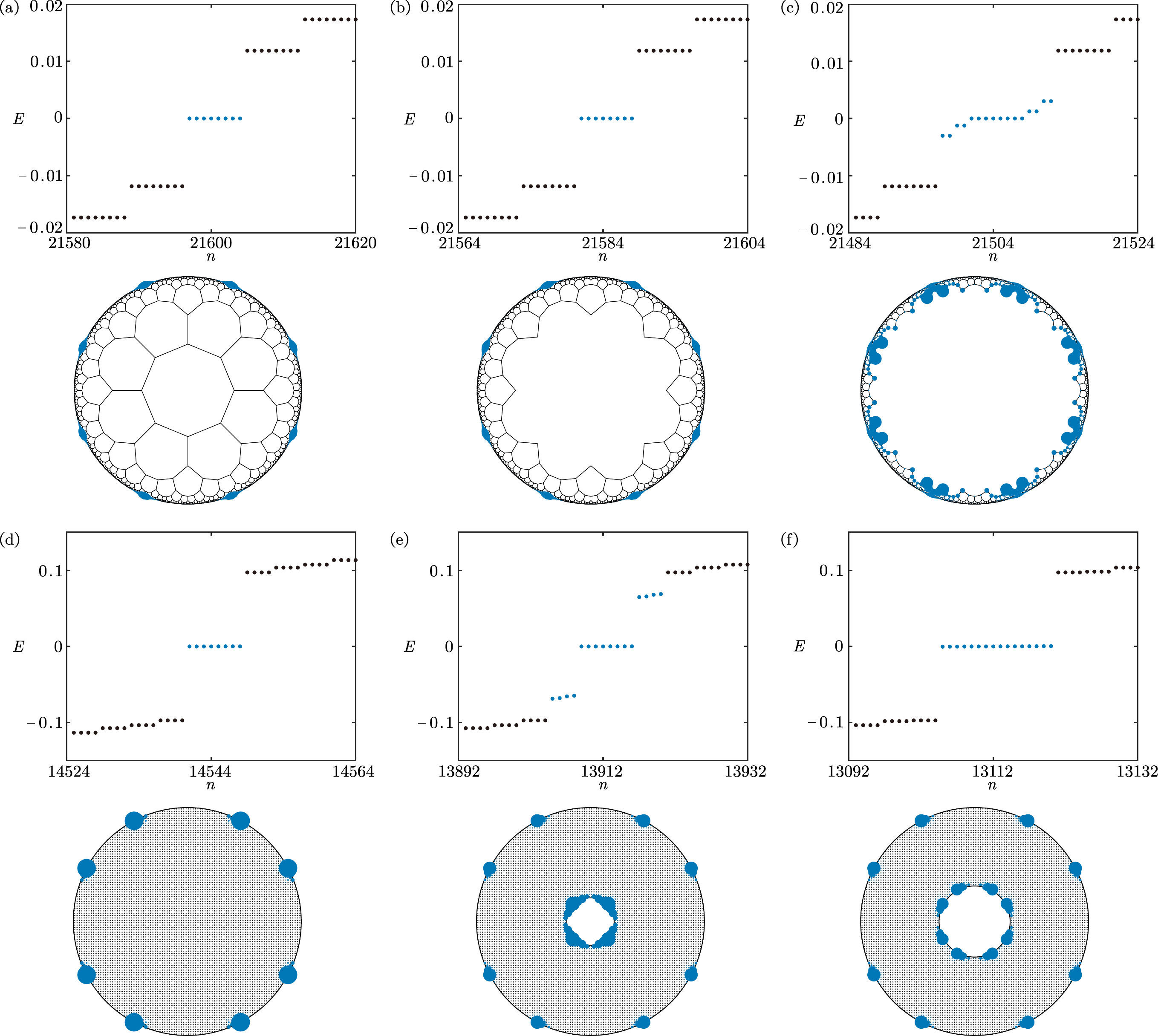}
\centering
\caption{The top panels of (a), (b), and (c) show the energy spectrum of the Hamiltonian $H$ for the type-I hyperbolic $\{8, 3\}$ lattice, the type-I $\{8, 3\}$ lattice with the innermost layer of sites removed, and the type-I $\{8, 3\}$ lattice with the innermost two layers of sites removed, respectively. The probability distributions of the eigenstates marked by blue dots are displayed in the bottom panels of (a), (b), and (c). In (a), (b), and (c), we set $M = -1.2$, $g = 0.5$, and $\eta = 4$, with the finite type-I hyperbolic $\{8, 3\}$ lattice consisting of six layers. The top panels of (d), (e), and (f) show the energy spectrum of the Hamiltonian $H$ for a Euclidean square lattice confined to a disk of radius $r = 48a$, to an annular geometry with $10a < r < 48a$, and to an annular geometry with $15a < r < 48a$, respectively. The probability distributions of the eigenstates marked by blue dots are displayed in the bottom panels of (d), (e), and (f). In (d), (e), and (f), we take $M = -2$, $g = 1$, and $\eta = 4$, with the lattice constant $a = 1$.}%
	\label{figC2}
\end{figure*}

Second, we introduce the modified BHZ model into type-I hyperbolic lattices with an artificially created central hole. As shown in Fig.~\ref{figC2}(a), when a finite type-I hyperbolic lattice has no central hole, eight zero-energy cornerlike states exist on the outer boundary. When a central hole of small radius is introduced into the type-I hyperbolic lattice, the zero-energy cornerlike states on the outer boundary remain robust, while no cornerlike states appear on the inner boundary, as shown in Fig.~\ref{figC2}(b). Furthermore, when a larger central hole is introduced, as shown in Fig.~\ref{figC2}(c), the zero-energy cornerlike states on the outer boundary remain stable, and cornerlike states emerge on the inner boundary but deviate from zero energy due to finite-size effects. Due to computational limitations, we are unable to construct type-I hyperbolic lattices with a larger number of layers. We anticipate that, given a sufficiently large number of layers, both the inner and outer boundaries of a type-I hyperbolic lattice with a central hole would host effective zero-energy cornerlike states.

Third, we introduce the modified BHZ model into Euclidean square lattices confined to an annular geometry. As illustrated in Figs.~\ref{figC2}(d)-(f), when the modified BHZ model is implemented in a Euclidean square lattice confined to an annular geometry, the cornerlike states gradually approach zero energy as the inner boundary radius increases, while the cornerlike states on the outer boundary remain robustly pinned at zero energy throughout this process.

In summary, we demonstrate that due to the equal number of sites on the inner and outer boundaries of type-II hyperbolic lattices, the cornerlike states on both boundaries respond identically to finite-size effects. By contrast, in type-I hyperbolic lattices with an introduced central hole and in square lattices confined to an annular geometry, the number of sites on the outer boundary substantially exceeds that on the inner boundary, leading to a distinct response of the cornerlike states on each boundary to finite-size effects.

%\newpage
\bibliographystyle{apsrev4-1-etal-title_6authors}

\end{document}